\pdfoutput=1

\documentclass[twocolumn,
aps,
reprint,
superscriptaddress,
]{revtex4-1}

\usepackage{graphicx}
\usepackage{bm}
\usepackage{amsmath}
\usepackage{gensymb}
\usepackage{hyperref}
\usepackage{amssymb}
\usepackage{pifont}
\usepackage{color}
\usepackage{upgreek}
\usepackage{tikz}

\begin{document}

\title{Evidence of standing spin-waves in a van der Waals magnetic material}
\author{Lucky N. Kapoor}
\thanks{these authors contributed equally}
\affiliation{Department of Condensed Matter Physics and Materials Science, Tata Institute of Fundamental Research, Homi Bhabha Road, Mumbai 400005, India}
\author{Supriya Mandal}
\thanks{these authors contributed equally}
\affiliation{Department of Condensed Matter Physics and Materials Science, Tata Institute of Fundamental Research, Homi Bhabha Road, Mumbai 400005, India}
\author{Meghan Patankar}
\affiliation{Department of Condensed Matter Physics and Materials Science, Tata Institute of Fundamental Research, Homi Bhabha Road, Mumbai 400005, India}
\author{Soham Manni}
\affiliation{Department of Condensed Matter Physics and Materials Science, Tata Institute of Fundamental Research, Homi Bhabha Road, Mumbai 400005, India}
\affiliation{Department of Physics, Indian Institute of Technology Palakkad, Palakkad 678557,India}
\author{A. Thamizhavel}
\affiliation{Department of Condensed Matter Physics and Materials Science, Tata Institute of Fundamental Research, Homi Bhabha Road, Mumbai 400005, India}
\author{Mandar M. Deshmukh}
\email{deshmukh@tifr.res.in}
\affiliation{Department of Condensed Matter Physics and Materials Science, Tata Institute of Fundamental Research, Homi Bhabha Road, Mumbai 400005, India}

\begin{abstract}
Spin-waves have been studied for data storage, communication and logic circuits in the field of spintronics based on their potential to substitute electrons\cite{chumak2017magnonic,chumak2015magnon}. Recent discovery of magnetism in two-dimensional (2D) systems such as monolayer CrI$_3$ and Cr$_2$Ge$_2$Te$_6$ has led to a renewed interest in such applications of magnetism in the 2D limit\cite{gong_discovery_2017,huang_layer-dependent_2017,gibertini_magnetic_2019}. Here we present direct evidence of standing spin-waves along with the uniform precessional resonance modes in van der Waals magnetic material, CrCl$_3$. Our experiment is the first direct observation of standing spin-wave modes, set up across a thickness of 20~$\bm{\upmu}$m, in a van der Waals material. We detect standing spin-waves in the vicinity of both, optical and acoustic, branches of the antiferromagnetic resonance. We also observe magnon-magnon coupling, softening of resonance modes with temperature and extract the evolution of interlayer exchange field as a function of temperature.
\end{abstract}

\maketitle

Spin-waves in various materials have shown potential for realization in applications such as wave-based computing and utilization in magnon-magnon interaction based nonlinear effects\cite{chumak2017magnonic,chumak2015magnon}. Materials with low magnetic damping and compatibility with micro-sized patterning are of particular interest. As spin-waves can also be generated by other methods such as spin-transfer torque\cite{lee2004excitations,madami2011direct}, the possibility of combining them with 2D materials can open up access to devices with multi-functional properties. Of particular interest is the family of chromium trihalides with a strong intralayer exchange, which enables long-range ferromagnetic ordering, and weak interlayer exchange, enabling realization of either ferromagnetic or antiferromagnetic ordering across layers\cite{crcl3-structure-mcguire2017crystal,cleavage-energy-theory-values-zhang2015robust}.
As antiferromagnets produce no stray fields due to their net zero magnetization and are generally robust against magnetic perturbations, they have a great potential for spintronics applications.

In this study, we have explored a van der Waals antiferromagnet chromium trichloride (CrCl$_3$) which has a very weak interlayer coupling of around 1.6 $\upmu$eV (ref. \cite{narath_spin_wave_analysis_1965}). The weak interlayer coupling allows us to study the spin dynamics and interactions in the system at microwave frequencies ($\sim$GHz), overcoming the challenges involved with the requirement of using THz radiation as in the case of other antiferromagnets\cite{terahertz-control-kampfrath2011coherent,terahertz-2-bhattacharjee2018neel}. CrCl$_3$ has also been reported to have low cleavage energy similar to the other chromium trihalides\cite{mcguire_crcl3_2017,mcguire-CrI3-2015,cleavage-energy-theory-values-zhang2015robust}. This opens up the possibility of  potential applications of few layer CrCl$_3$ with other 2D materials in van der Waals antiferromagnetic spintronics.

\begin{figure}[!]
\includegraphics{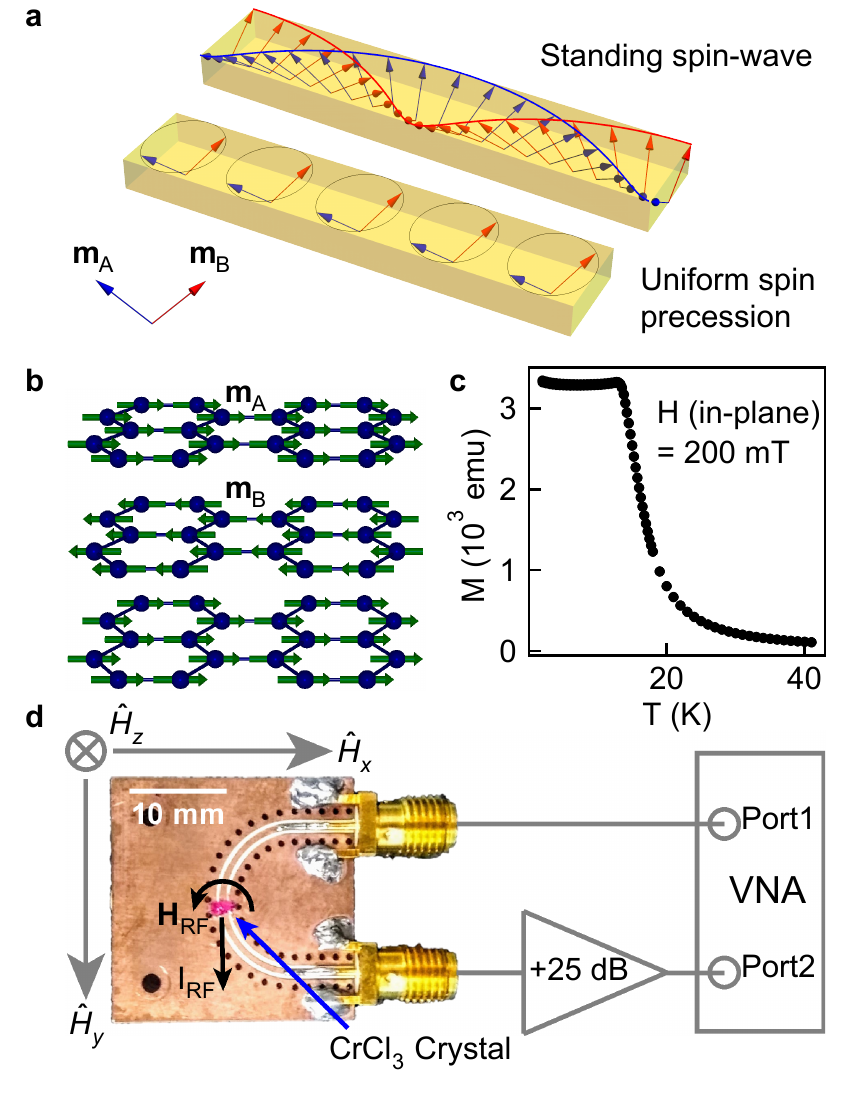}
\caption{ \label{fig:fig1} \textbf{Magnetization dynamics and crystal structure of CrCl$_3$, a van der Waals antiferromagnet.} (a) Schematic representation of uniform spin precession and standing spin-wave due to boundary conditions in an antiferromagnet. (b) Representation of the magnetic ordering of CrCl$_3$ crystal well below the ordering temperature, in absence of any applied magnetic field. The blue spheres represent the Cr$^{3+}$ ions and the green arrows represent the associated magnetic moments of the ions in the easy plane. (c) Magnetization vs temperature plot of a CrCl$_3$ crystal, with an applied in-plane static magnetic field of 200 mT, shows a change in magnetic ordering around 17 K\cite{mcguire_crcl3_2017,kuhlow_magnetic_1982}. (d) Image of a CrCl$_3$ crystal placed on a coplanar waveguide (CPW) used for this study along with the measurement scheme. Components of the static magnetic field ($\bm{\textbf{H}}_{\textrm{DC}}$) applied to the sample are along the unit vectors $\hat{H}_x$, $\hat{H}_y$ and $\hat{H}_z$.}
\end{figure}

We study the resonance modes of CrCl$_3$ as the spin dynamics is excited using microwave frequency signals in presence of magnetic field. On application of a static magnetic field, antiferromagnetic and ferromagnetic resonance modes can be observed depending on the magnetic structure; these are uniform spin precession modes where adjacent spins precess in sync, with a zero relative phase difference. These modes have been recently observed in CrCl$_3$ (ref. \cite{Macneill_GigahertzCrCl3_2019}). In addition to these, adjacent spins can precess with a nonzero relative phase difference, resulting in propagating spin-waves which are capable of carrying information. Due to the presence of boundary conditions, they form standing spin-wave modes. A schematic representation of these spin dynamics is shown in Fig.~\ref{fig:fig1}(a). The highlight of our work is the observation of these standing spin-wave modes in a van der Waals material, in addition to the antiferromagnetic and ferromagnetic resonance modes. We measure the temperature evolution of the resonance modes for different orientations of applied static magnetic field ($\bm{\textbf{H}}_{\textrm{DC}}$) relative to RF magnetic field ($\bm{\textbf{H}}_{\textrm{RF}}$) and the crystal easy plane. We also observe magnon-magnon coupling in CrCl$_3$ that has been reported recently\cite{Macneill_GigahertzCrCl3_2019}. Additionally, we observe the softening of the modes resulting from reduction in intralayer and interlayer exchange as the temperature is increased. We also measure reduction in the value of interlayer exchange field ($H_E$) with temperature.

\begin{figure*}[!]
\includegraphics{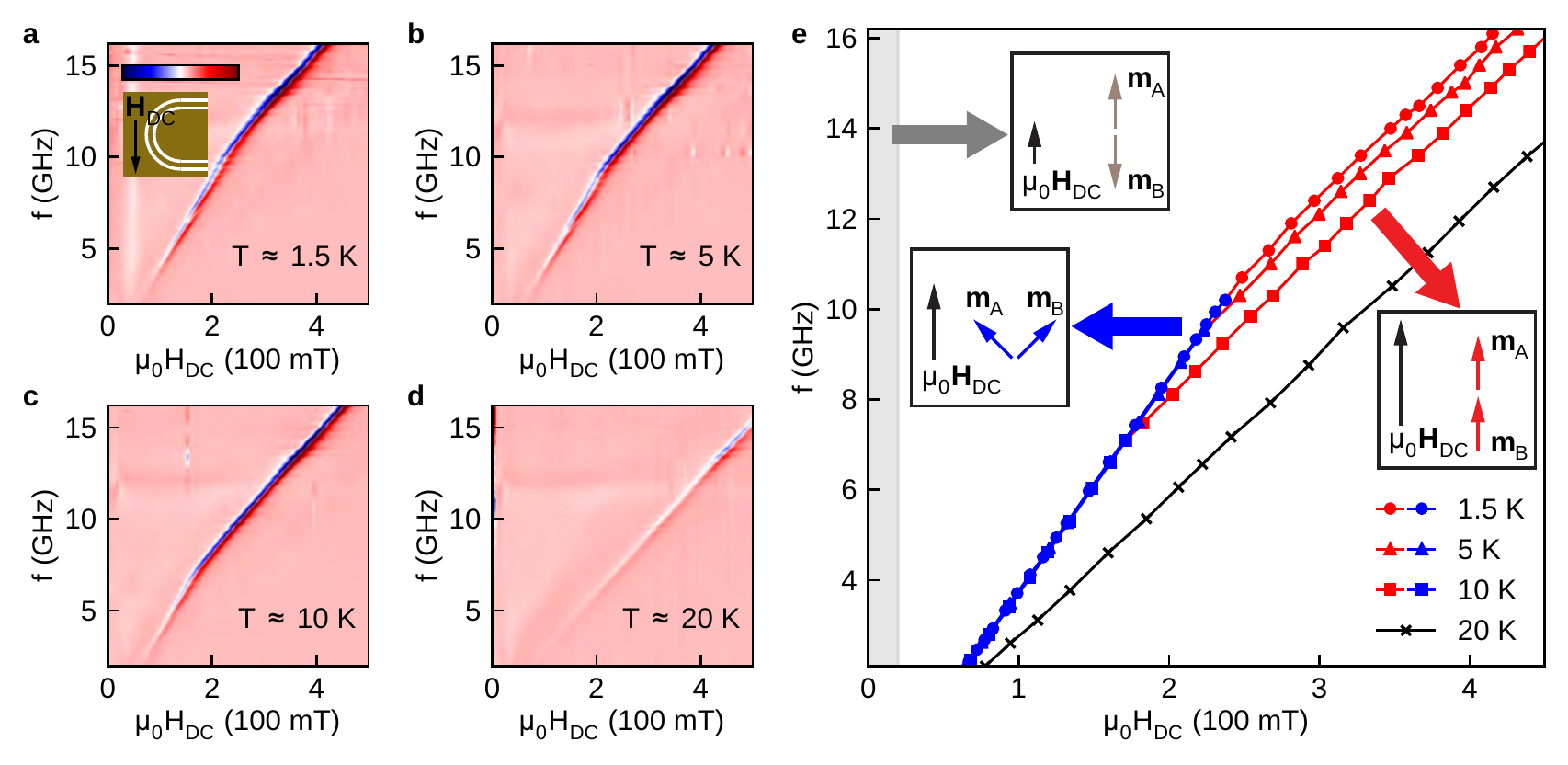}
\caption{\label{fig:fig2} \textbf{Probing orientation of magnetic moments using antiferromagnetic resonance.} (a), (b), (c) and (d) Color-scale plot showing derivative of magnitude of transmission coefficient, $|S_{21}|^2$ with respect to $H_{\textrm{DC}}$, $\frac{1}{\mu_0}\frac{d}{dH_{\textrm{DC}}}|S_{21}|^2$, as a function of microwave frequency and $\bm{\textbf{H}}_{\textrm{DC}}$ applied along $\hat{H}_y$ direction, at temperatures 1.5~K, 5~K, 10~K and 20~K respectively. Colorscale for these data (shown in the inset of (a)) ranges linearly from -45 dB/T to 35 dB/T. (e) Comparison of the dispersion in (a)-(d). In (a)-(c), there is a change in slope of the resonance feature at higher magnetic fields which is highlighted in (e). Gray refers to the region before spin-flop ($\mu_0H_{\textrm{DC}} < 15\ \textrm{mT}$; ref. \cite{kuhlow_magnetic_1982}); blue shows the spin orientation for canted ordering state (corresponding to acoustic antiferromagnetic resonance); red shows the spin orientation after the change in slope of the resonance modes in the parallel ordering state (corresponding to ferromagnetic resonance). The black curve shows the electron spin resonance at 20 K.}
\end{figure*}

\begin{figure*}[!]
\center
\includegraphics{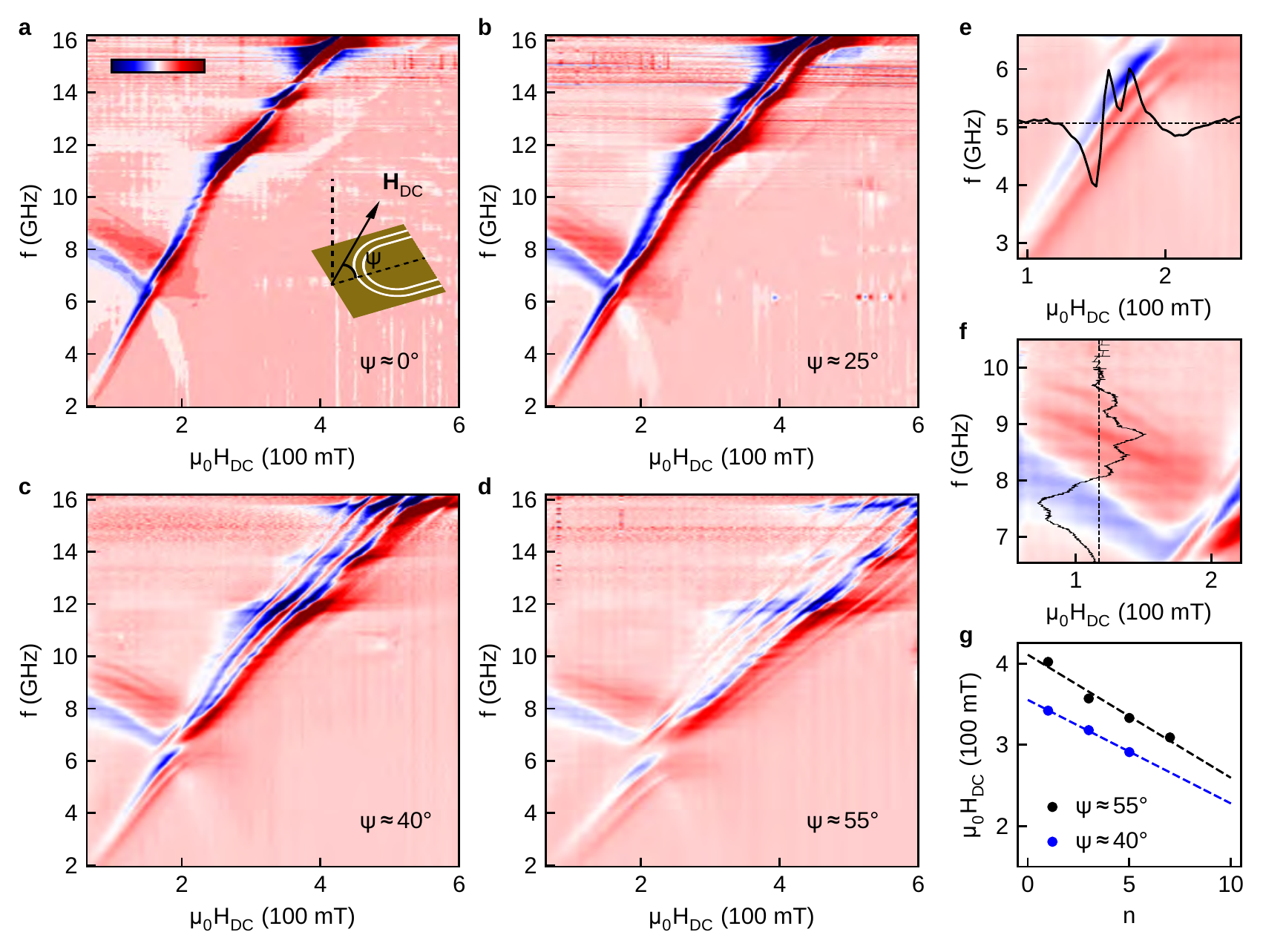}
\caption{\label{fig:fig3} \textbf{Standing spin-wave modes in CrCl$_3$.} (a), (b), (c) and (d) Color-scale plot showing derivative of magnitude of transmission coefficient, $\frac{1}{\mu_0}\frac{d}{dH_{\textrm{DC}}}|S_{21}|^2$, with respect to $H_\textrm{DC}$, as a function of microwave frequency and $H_\textrm{DC}$ as the out-of-plane component of $\bm{\textbf{H}}_\textrm{DC}$ (applied in \emph{xz}-plane) is increased by changing the tilt angle of the CPW with respect to the \emph{xy}-plane. Colorscale for these data (shown in the inset of (a)) ranges linearly from -24 dB/T to 20 dB/T. The multiplicity of the modes increases as the angle $\psi$ is increased. Data has been shown for  measurements done at $T\approx1.5~$K. (e) and (f) show the resonance features in the antiferromagnetic regime corresponding to acoustic and optical modes respectively for $\psi \approx 55^\circ$ . The additional peaks can also be seen in the overlaid line plots. (g) shows the linear dependence of the position of the spin-wave modes with odd integers, $n$, for $f = 11~$GHz, according to equation~(\ref{eq:1}).}
\end{figure*}

For the study of magnetic structure of CrCl$_3$, the crystal is grown using vapor transport method used earlier\cite{mcguire_crcl3_2017}. Below its N\'eel temperature\cite{crcl3-neel-temperature-cable1961neutron}, in the absence of magnetic field, CrCl$_3$ has a magnetic structure\cite{crcl3-structure-narath1963low,crcl3-structure-mcguire2017crystal,mcguire_crcl3_2017} as shown in Fig.~\ref{fig:fig1}(b). Cr$^{3+}$ ions provide the localized magnetic moments in the lattice. Within a layer, there is ferromagnetic coupling between the magnetic moments while adjacent layers have antiferromagnetic coupling. Overall the system possesses an easy plane anisotropy. Fig.~\ref{fig:fig1}(c) shows the change in magnetization of the system close to the ordering temperature, similar to previously reported values\cite{mcguire_crcl3_2017,kuhlow_magnetic_1982}. Fig.~\ref{fig:fig1}(d) shows a small flake (thickness $\approx$ 20~$\upmu$m) of CrCl$_3$ placed on a coplanar waveguide (CPW) used in this study. The crystal is placed such that the easy plane is parallel to the plane of the CPW. The sample is loaded in a cryostat in different orientations with respect to the direction of applied static magnetic field, $\bm{\textbf{H}}_\textrm{DC}$, (as shown in Fig.~\ref{fig:fig1}(d)) and cooled by a continuous flow of helium vapor around the sample. We measure the transmission coefficient, $S_{21}$, as a function of frequency ($f$) for different $H_\textrm{DC}$ values. Methods gives details about the crystal growth as well as measurement setup.

When $\bm{\textbf{H}}_\textrm{DC}$ is applied parallel to the plane of the crystal and perpendicular to $\bm{\textbf{H}}_\textrm{RF}$ (along the $\hat{H}_y$ direction), a linearly dispersing resonance mode is observed as shown is Fig.~\ref{fig:fig2}(a-d). To understand this mode, one can use the macrospin approximation for CrCl$_3$ proposed by MacNeill \emph{et al.}\cite{Macneill_GigahertzCrCl3_2019} that is similar to the description of Liensberger \emph{et al.}\cite{liensberger_exchange-enhanced_2019}. As there are two sublattices in the magnetic unit cell, there will be two primary resonance modes. When $\bm{\textbf{H}}_{\textrm{DC}}$ is applied parallel to the plane of the crystal and along the direction of the RF current, only the odd mode is excited as in-plane and out-of-plane components of $\bm{\textbf{H}}_\textrm{RF}$ are both odd under twofold rotation around $\hat{H}_y$. This is analogous to the ``acoustic" mode in weakly coupled mechanical oscillators and the magnetization oscillations of the two sublattices are in phase. However, when $\bm{\textbf{H}}_\textrm{DC}$ is along $\hat{H}_x$, the in-plane component of $\bm{\textbf{H}}_\textrm{RF}$ is even and the out-of-plane component of $\bm{\textbf{H}}_\textrm{RF}$ is odd under twofold rotation around $\hat{H}_x$; thus both the modes are excited. This mode is analogous to the ``optical" mode in weakly coupled mechanical oscillators. For $\mu_0H_{\textrm{DC,}z} = 0$~T, there is a degeneracy at the point where the two modes cross each other as the two modes are protected by the symmetry of the anisotropy energy about $\bm{\textbf{H}}_\textrm{DC}$ (ref.\cite{liensberger_exchange-enhanced_2019}). When the static magnetic field has an out-of-plane component, this symmetry is broken and there is a hybridization of the modes leading to an avoided crossing.

In order to understand the microscopic magnetization dynamics, specific temperature slices have been shown in Fig.~\ref{fig:fig2}(a-d) for $\bm{\textbf{H}}_{\textrm{DC}}$ applied in-plane along $\hat{H}_y$ direction. For $\mu_0H_{\textrm{DC}} < 15$~mT (ref. \cite{kuhlow_magnetic_1982}), the two sublattices are anti-parallel in orientation with no net magnetic moment over the magnetic unit cell as shown by the gray colored spins in Fig.~\ref{fig:fig2}(e). This feature has been attributed to magnetic reorientation of the spins, known as spin flop.  In bulk CrCl$_3$ crystals, some indication of spin flop has been previously observed  around 10-20 mT (ref. \cite{kuhlow_magnetic_1982,mcguire_crcl3_2017,narath_spin_wave_analysis_1965}). When an in-plane magnetic field is applied, spin flop transition occurs in the moments, as shown by the blue colored spins in Fig.~\ref{fig:fig2}(e). The spins get canted with respect to the field direction and consequently couple to the magnetic field. Here, as $\bm{\textbf{H}}_{\textrm{DC}}$ has been applied along the direction of the RF current, only the acoustic mode is excited for the canted spins.

The resonance mode changes slope around $\mu_0 H_{\textrm{DC}} \approx 250$~mT, due to the change in magnetic ordering from a canted state to parallel orientation of the two sublattices as seen in Fig.~\ref{fig:fig2}(a-c). The resonance associated with the moments in parallel orientation of the sublattices is akin to the ferromagnetic resonance mode. Beyond 250~mT, the resonance mode satisfies the uniform ferromagnetic resonance condition given by $2 \pi f = \mu_0 \gamma \sqrt{H_{\textrm{DC}} (H_{\textrm{DC}} + M_S)}$  where $\gamma$ is the gyromagnetic ratio and $M_S$ is the saturation magnetization representing the contribution of anisotropy energy. As the temperature is increased, the interlayer exchange decreases and hence the transition to ferromagnetic state happens at $\mu_0 H_{\textrm{DC}}$ $<$ 250 mT. Beyond the ordering temperature $\approx$ 17 K, a linear resonance mode with slope 30.62 GHz/T ($g$-factor = 2.18) is seen in Fig.~\ref{fig:fig2}(d) corresponding to the electron paramagnetic resonance of Cr$^{3+}$ ions\cite{chehab1991two}.  Fig.~\ref{fig:fig2}(e) shows the evolution of the resonance modes at different temperatures and one can clearly observe the reduction in the field, for the transition from antiferromagnetic to ferromagnetic sublattice structure. Spin orientation and slope change have been highlighted with blue and red color corresponding to the canted and parallel ordering respectively.

\textbf{} When the magnetic field is applied in the $\hat{H}_x$ direction, we see both optical and acoustic modes together (as shown in Supplementary section III). However, on introducing an out-of-plane component of $\bm{\textbf{H}}_{\textrm{DC}}$ as well, two additional features are observed: avoided crossing between the two modes due to magnon-magnon interaction and excitation of some additional resonance modes. Fig.~\ref{fig:fig3} shows these features for the antiferromagnetic resonance of CrCl$_3$ as a function of angle of $\bm{\textbf{H}}_{\textrm{DC}}$ relative to the plane of the CPW, $\psi$. Multiplicity of modes is seen for both antiferromagnetic and ferromagnetic ordering of the sublattices. Of particular interest is the observation of the multiplicity of resonant modes for both optical and acoustic branches of the antiferromagnetic resonance. We also note that the number of resonance features in the vicinity of the acoustic mode and the ferromagnetic resonance increases as the angle of $\bm{\textbf{H}}_{\textrm{DC}}$ deviates from the plane of the film.

\begin{figure}[b]
\includegraphics{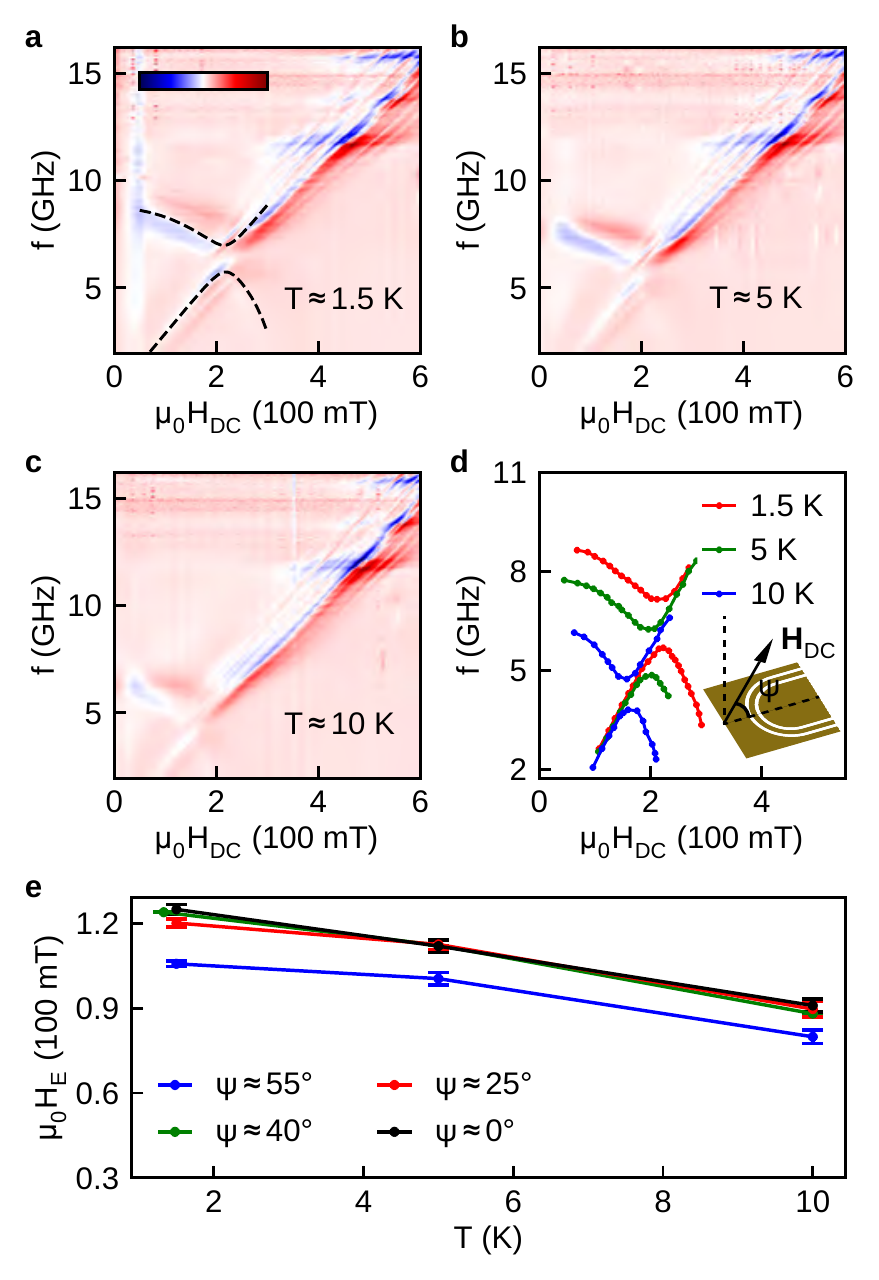}
\caption{\label{fig:fig4} \textbf{Tuning magnon-magnon coupling in CrCl$_3$.} (a), (b) and (c) Colorscale plot of derivative of magnitude of transmission coefficient, $\frac{1}{\mu_0}\frac{d}{dH_{\textrm{DC}}}|S_{21}|^2$, with respect to $H_\textrm{DC}$ as a function of microwave frequency and $H_\textrm{DC}$ with $\bm{\textbf{H}}_\textrm{DC}$ applied out-of-plane at an angle of 55$^\circ$ with respect to the \emph{xy}-plane and perpendicular to the RF current, show the evolution of the hybridized modes  due to the out-of-plane component at temperatures 1.5 K, 5 K and 10 K respectively. Colorscale for these data (shown in the inset of (a)) ranges linearly from -33 dB/T to 30 dB/T. (a) also shows an overlay of data fitted to the model presented in ref. \cite{Macneill_GigahertzCrCl3_2019}. (d) Softening of the hybridized modes with increase in temperature observed by overlaying the modes measured as a function of temperature. (e) Interlayer exchange field, $H_E$, calculated at different temperatures from the fitting of the curves for two modes to the mentioned model; the values have been derived for magnetic field applied at $\psi \approx 0^\circ$, 25$^\circ$, 40$^\circ$ and 55$^\circ$ (more details in Supplementary section IV).}
\end{figure}

The microscopic origin of the additional resonances that we observe can be understood as standing spin-wave modes being set up across the thickness of the crystalline film. The idea of spin-wave resonances in the context of the ferromagnetic films was proposed \cite{kittel_excitation_1958} and it was then experimentally observed in ferromagnets\cite{seavey_direct_1958} and antiferromagnets\cite{lui1990antiferromagnetic}. The standing spin-wave modes have been studied widely and the surface anisotropy \cite{puszkarski_theory_1979} plays a key role in determining the symmetry of the spin-wave modes that are set up along the thickness of the crystal and their angular dependence\cite{phillips_spin_1966}. For a uniformly magnetized crystalline film, the Kittel model\cite{kittel_excitation_1958} proposes a dependence of the spectrum such that the position of the $n$th spin-wave resonance mode is given by
$H_n = H_0 -n^2\frac{D}{g\mu_B}\frac{\pi^2}{L^2}$ where $H_0$ is the position of the ferromagnetic mode corresponding to uniform spin precession, $D$ is the exchange stiffness constant, $L$ is the thickness of the sample and $n$ is an odd integer. Thus, the position of the spin-wave resonance mode will show a linear dependence in $n^2$. However, as shown in Fig.~\ref{fig:fig3}(g), the spin-wave modes seen in CrCl$_3$ show a linear dependence in $n$. This can be explained by the model proposed by Portis\cite{portis1963low} for a non-uniform volume magnetization within the film, assuming a symmetrical parabolic drop from the center of the film to the edges. Then the position of the $n$th spin-wave resonance mode is given by\cite{liu2007angular}
\begin{equation}\label{eq:1}
    H_n = H_0 - \Big( n - \frac{1}{2} \Big) (4/L) \Big( 4\pi {M^{0}}_{\textrm{eff}}\varepsilon\frac{D}{g\mu_B}\Big)^{\frac{1}{2}}
    \end{equation}
where $4\pi{M^{0}}_\textrm{eff}$ is the magnetization in the center and $\varepsilon$ is the distortion parameter for the crystalline film. Fig.~\ref{fig:fig3}(g) shows this linear dependence of spin-wave resonance modes with $n$ for the case of $\psi \approx 40^\circ$ and $\psi \approx 55^\circ$. The non-uniform magnetization suggests possible changes in previously presented models to accurately describe this system. We also see signature of even modes which is possible for inhomogeneous magnetic field (shown in Supplementary section V). We observe in Fig.~\ref{fig:fig3} (e) and (f) that the multiplicity of modes is present even when the system has an antiferromagnetic state, seen in both the acoustic and optical modes respectively. This is the first direct observation of spin-wave resonance modes due to geometric boundary conditions in a van der Waals magnetic material.

In Fig.~\ref{fig:fig3}(b-d), an avoided crossing is also seen when $\bm{\textbf{H}}_{\textrm{DC}}$ has an out-of-plane component; this is due to magnon-magnon coupling. From the evolution of the resonance modes, we can extract the interlayer exchange field as a function of temperature. As the temperature is increased, there is a softening of the hybridized modes and the avoided crossing occurs at a lower frequency and magnetic field.

A quantitative description of the hybridized modes has been presented in ref. \cite{Macneill_GigahertzCrCl3_2019}. While solving the LLG equation we get two modes -- the acoustic mode with resonance frequency, $f_a  = \mu_0 \frac{\gamma}{2\pi} \big( {1+ \frac{M_S}{2H_E}}\big) ^{1/2} H \cos \psi $ and the optical mode with resonance frequency, $f_o = \mu_0 \frac{\gamma}{2\pi} \big({2H_E M_S (1- \frac{H^2}{{H_{\textrm{FM}}}^2}) + \frac{{\sin}^2 \psi}{(1+ (M_S/2H_E))^2}H^2}\big) ^{1/2}$. $H_{\textrm{FM}}$ is the magnetic field required to align both the sublattices and it is given by $ 1/{H_{\textrm{FM}}}^2 = \cos^2 \psi / {(2H_E)}^2 + \sin^2 \psi / {(2H_E + M_S)}^2$.




We study the temperature evolution of modes as shown in Fig.~\ref{fig:fig4}(d) and fit the evolution of modes as a function of magnetic field to the experimental data; this allows us to determine $H_E$ as a function of temperature. Fig.~\ref{fig:fig4}(a) shows the simultaneous fitting (details of the fitting procedure provided in Supplementary section IV) of the two branches for $\psi \approx 55 ^\circ$ and T $\approx 1.5~$K. The magnitude of exchange field agrees with previous observations in antiferromagnetic systems \cite{Macneill_GigahertzCrCl3_2019,kuhlow_magnetic_1982,narath1964nuclear}. In addition,  a decreasing trend is seen in the interlayer exchange field as a function of temperature similar to other systems\cite{temp-dependence-persat1997strong,wang_temperature_2018}. As the temperature is increased from 1.5~K to 10~K, the interlayer exchange field reduces by $\approx 30\%$.

We also extract lifetime of magnons corresponding to the acoustic and optical branches of the antiferromagnetic resonance as well as for the ferromagnetic resonance by fitting a Lorentzian to the $|S_{21}|^2$ data as a function of frequency. For the case of $\bm{\textbf{H}}_\textrm{DC}$ parallel to the $\hat{H}_x$ direction, the lifetimes of magnons for the optical and acoustic modes of the antiferromagnetic resonance are around 2.4 ns and 3.0 ns at low fields. For the ferromagnetic resonance, the lifetime is around 5.0 ns. Additionally, we extract Gilbert damping constant for the ferromagnetic mode by fitting a Lorentzian to the $|S_{21}|^2$ data as a function of the static magnetic field. From a linear fit of the ferromagnetic resonance linewidths ($\Delta H$) at different frequencies, we report the Gilbert damping constant to be around $2.0 \times 10^{-3}$; an order of magnitude higher than YIG\cite{dubs2017sub} (more details of the fitting procedures in Supplementary section VI).

Our work presents a detailed study of the resonance modes of CrCl$_3$. For the first time, we observe standing spin-wave modes in a van der Waals material. These spin-wave modes, which can carry information without the movement of electrons, can be used as data carriers in low-loss systems. Of particular interest is that we detect standing spin-wave modes for both the branches of the antiferromagentic resonance. We observe a reduction of 30$\%$ in the interlayer exchange field with change in temperature from 1.5 K to 10 K. We also report values for the magnon lifetimes and Gilbert damping constant in CrCl$_{3}$. While this study was done for a bulk crystal of CrCl$_3$, it presents exciting prospects in the study of few layers CrCl$_3$ and its application in antiferromagnetic spintronics using van der Waals architecture.

We thank Akashdeep Kamra, Rajamani Vijayraghavan, Anjan Barman, P. L. Paulose, Ganesh Jangam, Atul Raut and Moumita Nandi for helpful discussion and experimental assistance.  We acknowledge the Swarnajayanti Fellowship of the Department of Science and Technology (for M.M.D.), DST Nanomission grant SR/NM/NS-45/2016, ONRG grant N62909-18-1-2058, and the Department of Atomic Energy of the Government of India for support.

\section*{Methods}
\subsection*{Crystal growth}
CrCl$_3$ single crystals were grown by chemical vapor transport (CVT) method from CrCl$_3$.6H$_2$O, chromium trichloride hexahydrate. CrCl$_3$.6H$_2$O powder (purity 99.5\%) was transformed into anhydrous CrCl$_3$ powder by heating at 90$^\circ$C while continuously pumping inside a horizontal tube furnace. Color of the powder changes from green to magenta after it loses its six water molecules. This anhydrous powder is immediately sealed inside an evacuated quartz tube. The sealed quartz tube is put into a three zone gradient tube furnace for CVT with hot zone at 700$^\circ$C and growth zone at 550$^\circ$C. After 15 days of growth, CrCl$_3$ crystals of 10-20 mm$^2$ dimension were obtained near the growth zone of the sealed tube.
\subsection*{Device fabrication and measurement}
The CrCl$_3$ crystal is transferred to a CPW in a glove box environment to avoid degradation of the crystal. The CPW is designed to have 50~$\Omega$ characteristic impedance, with central line width of 0.74~mm and gap of 0.3~mm. The crystal is covered by Apeizon N thermal grease to secure it on the CPW, to protect it from the ambient and to ensure that it has good thermal contact with the CPW. A temperature sensor (Lakeshore: DT470) is connected in direct contact with one of the corners of the CPW PCB. The CPW is connected to a vector network analyzer (Anritsu: MS46122B) using microwave transmission cables. The response of the system is captured after amplifying the transmission signal through a $\approx$ 25 dB amplifier (MiniCircuits: ZVA-213-S+). Same CrCl$_3$ crystal has been used for all measurements. Device characterization is done at approximate power level of -22 dBm at the sample for 6~GHz. Typical frequency dependent transmission, $|S_{21}|^2$, for zero field has been shown in Supplementary section I. We also record some field dependent background signals separately by measuring the response of the CPW PCB in absence of any CrCl$_3$ crystal. This data has been shown in Supplementary section II. Each plot has been obtained by combining slices of the field derivative of measured transmission coefficient, $\frac{1}{\mu_0}\frac{d}{dH_{\textrm{DC}}}|S_{21}|^2$, as a function of frequency for different $H_{\textrm{DC}}$ values.

\bibliography{References}

\onecolumngrid

\section{$\boldsymbol{|S_{21}|^2}$ data for different configurations}
\begin{figure*}[h]
\includegraphics{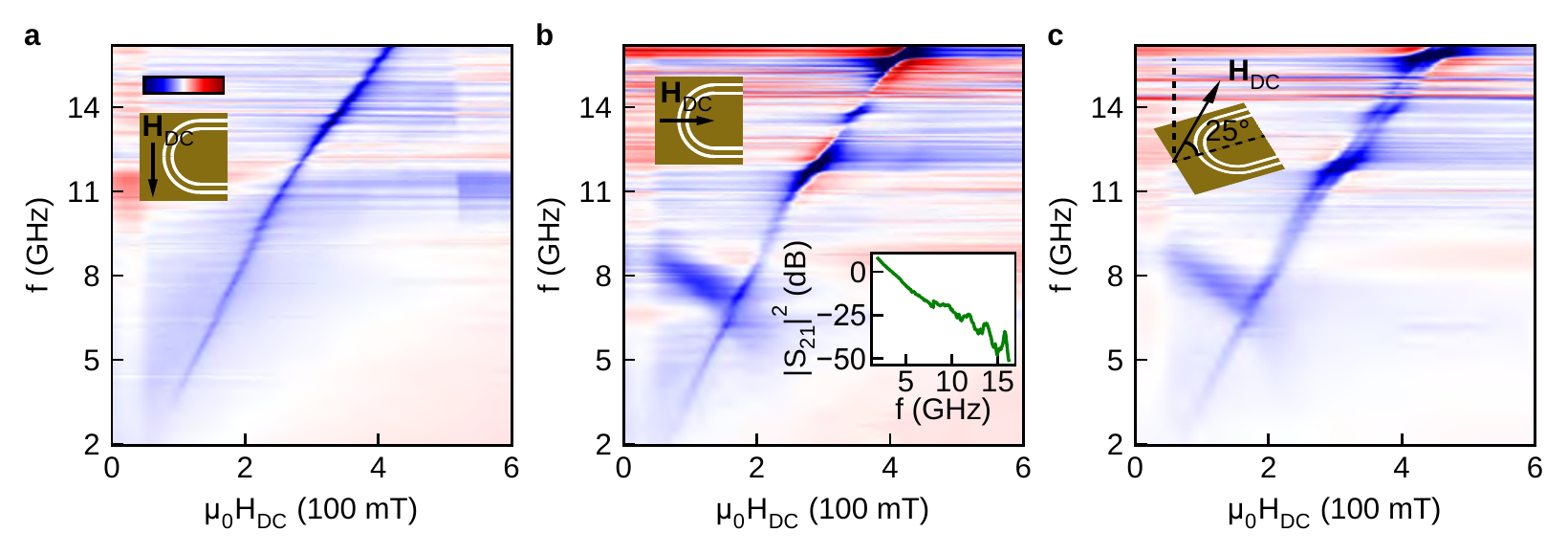}
\caption{\label{fig:fig5} \textbf{Color-scale plot of $\boldsymbol{|S_{21}|^2}$}. (a), (b) and (c) show the color-scale plot of the magnitude of transmission coefficient, $|S_{21}|^2$ (in dB),  as a function of microwave frequency and $H_{\textrm{DC}}$. (a) corresponds to the orientation of $\mathbf{H}_{\textrm{DC}}$ along $\hat{H}_y$ direction of the coordinate system defined in the Fig.~1 of main text. (b) corresponds to the orientation of $\mathbf{H}_{\textrm{DC}}$ along $\hat{H}_x$ direction and (c) corresponds to the orientation of $\mathbf{H}_{\textrm{DC}}$ at an angle of 25$^\circ$ from the CPW in the \emph{xz}-plane. All the measurement data shown correspond to $T \approx 1.5~$K.}
\end{figure*}

In the main text, color-scale plots of derivative of magnitude of transmission coefficient, $\frac{1}{\mu_0}\frac{d}{dH_{\textrm{DC}}}|S_{21}|^2$, have been used as the background of the transmission signal has frequency and temperature dependence. Fig.~\ref{fig:fig5} shows the the color-scale plot of the magnitude of transmission coefficient, $|S_{21}|^2$ (in dB). The plot is obtained by combining the measured transmission coefficient as a function of frequency for different $H_{\textrm{DC}}$ values. Each frequency slice was then background corrected by subtracting the average transmission coefficient over two different ranges of 0-21 mT and 579-600 mT corresponding to regions where no resonance is seen in the measured frequency range. In Fig.~\ref{fig:fig5}(a), the resonance feature corresponding to the acoustic mode is clearly visible and in Fig.~\ref{fig:fig5}(b), both, optical and acoustic modes are seen along with the point of degeneracy. In Fig.~\ref{fig:fig5}(c), we can also see the standing spin-wave resonance modes in the region corresponding to the parallel orientation of the two sublattices on application of a component of $\mathbf{H}_\textrm{DC}$ along $\hat{H}_z$ as well. However, a simple subtraction of data does not remove the background dependence of the transmission signal completely. Hence, we have used field derivative of $|S_{21}|^2$ for background elimination\cite{maier2018note}.

\section{Magnetic field and frequency dependence of background signal}
\begin{figure*}[h]
\includegraphics{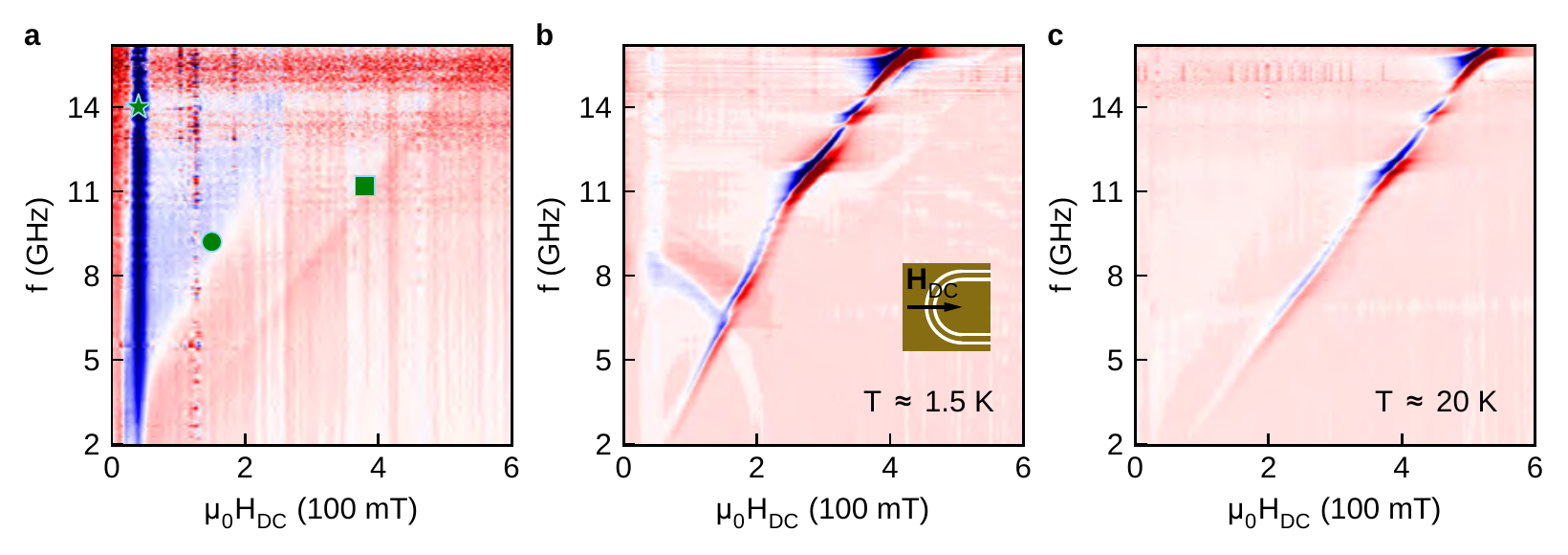}
\caption{\label{fig:fig6} \textbf{Background signal for the CPW.} (a), (b) and (c) show the color-scale plot of the magnitude of $|S_{21}|^2$ (in dB)  as a function of microwave frequency and $H_{\textrm{DC}}$. (a) shows the background signal for the CPW without any CrCl$_3$ crystal. (b) and (c) correspond to the orientation of $\mathbf{H}_{\textrm{DC}}$ along $\hat{H}_x$ direction for $T \approx 1.5$~K and $T \approx 20$~K respectively. Some features seen in (b) and (c), as well as other colorplots in the main text, can be attributed to the background signal shown in (a). These features have been marked using different symbols.}
\end{figure*}

Use of field derivative of $|S_{21}|^2$ removes the frequency dependent background signal in the data. However, there are other background signals which have dependence on magnetic field as well. Fig.~\ref{fig:fig6}(a) shows the field derivative of $|S_{21}|^2$  as a function of microwave frequency and $\mathbf{H}_{\textrm{DC}}$ applied along $\hat{H}_x$ for a bare CPW without any CrCl$_3$ crystal. There are three main features seen in this background signal 2D plot. A vertical feature is seen around $\mu_0 H_{\textrm{DC}} \approx 30~$mT, marked by $\bigstar$. The feature is also seen in Fig.~\ref{fig:fig6}(b) as well as some of the plots in the main text and other sections of the supplemental information. This feature persists upto field close to the spin-flop transition in CrCl$_3$. Such a strong background feature makes it difficult to detect any signal related to spin-flop transition, if it is detectable in such a resonance technique. Another feature, marked by $\blacksquare$, with linear dependence with $\mu_0 H_{\textrm{DC}}$ is also seen, with a slope similar to the ESR of the Cr$^{3+}$ ions in CrCl$_3$ (seen in Fig.~\ref{fig:fig6}(c)) and might be the ESR signal of Cu$^{2+}$ ions in the CPW PCB. Another feature, marked by \tikz\draw[black,fill=black] (0,0) circle (.5ex);, is also seen.

\section{Crossing of acoustic and optical mode}
\begin{figure*}[h]
\includegraphics{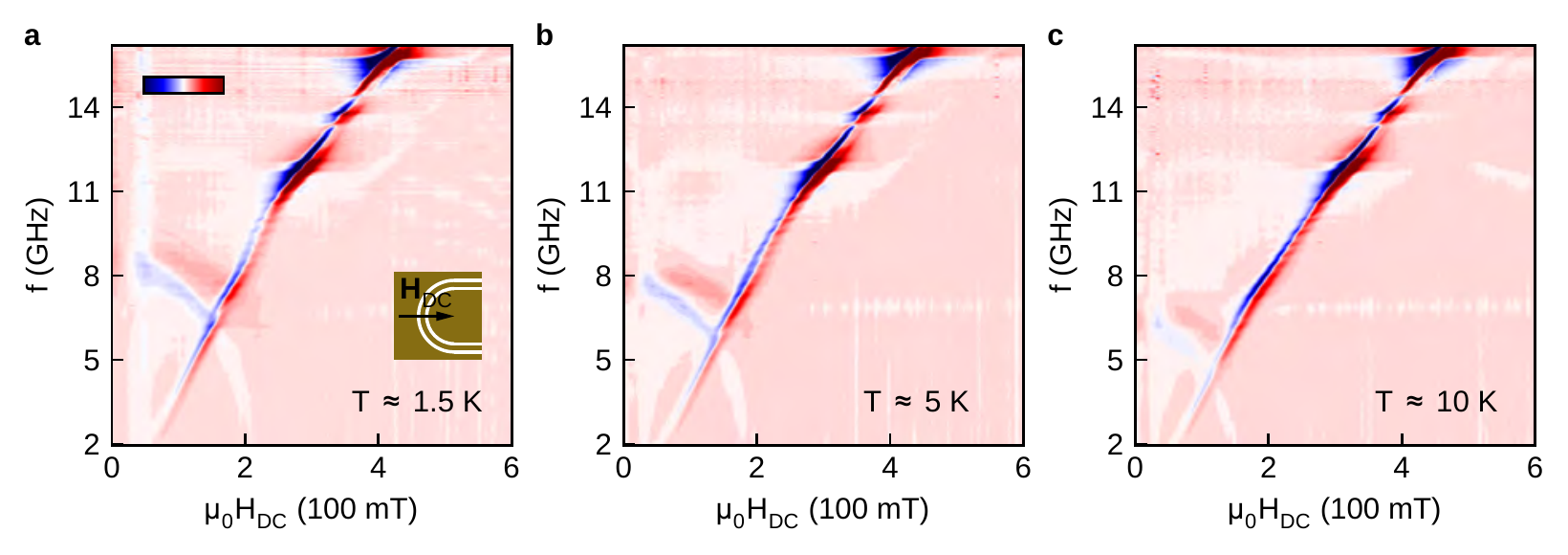}
\caption{\label{fig:fig7} \textbf{Degeneracy of modes for $H_{\textrm{DC},z}=0$~T.} (a), (b) and (c) Color-scale plot showing the derivative of magnitude of transmission coefficient, $\frac{1}{\mu_0}\frac{d}{dH_{\textrm{DC}}}|S_{21}|^2$, with respect to $H_\textrm{DC}$, as a function of microwave frequency and $H_\textrm{DC}$ for $\mathbf{H}_\textrm{DC}$ applied along $\hat{H}_x$ direction, at temperatures 1.5~K, 5~K and 10~K respectively. Degeneracy can be seen at the point where the two modes cross each other as the modes are protected by the symmetry of the anisotropy energy about $\mathbf{H}_\textrm{DC}$. Comparison of these plots show a decrease in resonance frequency (softening) of the modes as the temperature is increased.}
\end{figure*}

When $\mathbf{H}_\textrm{DC}$ is applied in plane and parallel to $\hat{H}_x$, both optical and acoustic resonance modes are excited leading to a degeneracy at the point where the two modes cross each other. This degeneracy can be lifted by applying a magnetic field in the \emph{xz}-plane with $H_{\textrm{DC},z}\neq0~$T. Fig.~\ref{fig:fig7} shows the color-scale plot of $\frac{1}{\mu_0}\frac{d}{dH_{\textrm{DC}}}|S_{21}|^2$ as a function of microwave frequency and $H_\textrm{DC}$ at different temperatures. Softening of the modes is also seen as the temperature increases.

\section{Simultaneous curve fitting to extract parameters}
\begin{figure*}[h]
\includegraphics{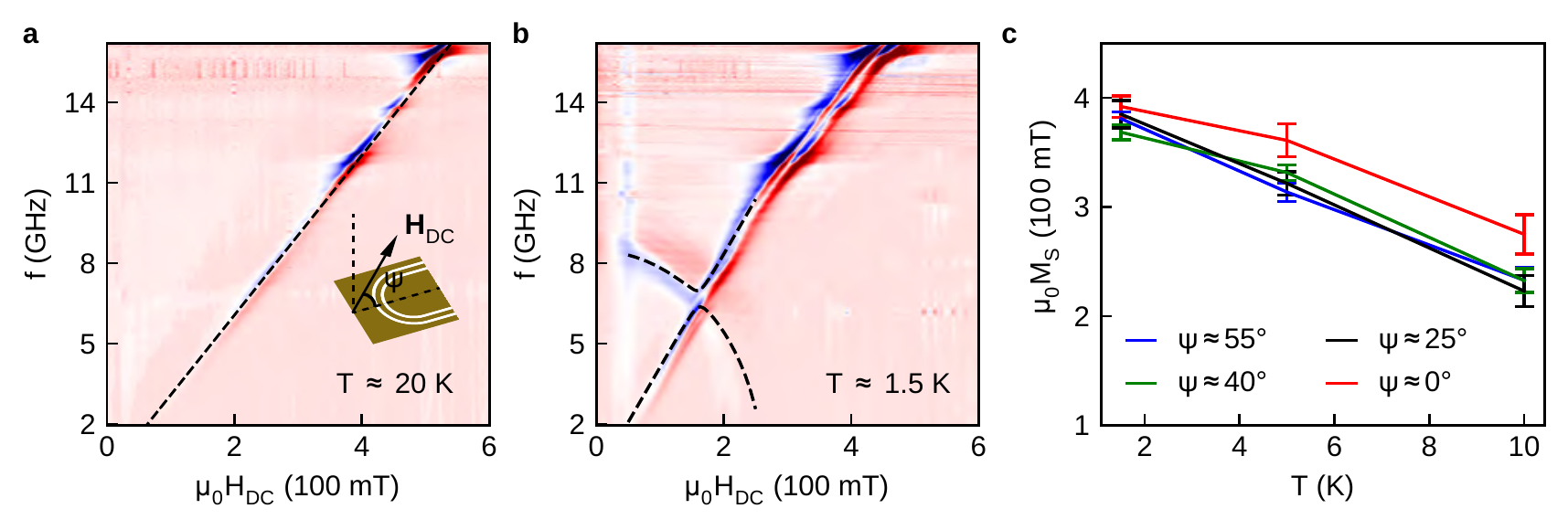}
\caption{\label{fig:fig8} \textbf{Simultaneous fitting of resonance modes.} (a) and (b) Color-scale plot showing the derivative of magnitude of transmission coefficient, $\frac{1}{\mu_0}\frac{d}{dH_{\textrm{DC}}}|S_{21}|^2$, with respect to $H_\textrm{DC}$, as a function of microwave frequency and $H_\textrm{DC}$ for $\mathbf{H}_\textrm{DC}$ applied at an angle $\psi \approx 25^\circ$ and perpendicular to the RF current, at temperatures 20~K, and 1.5~K respectively. Overlaid plot in (a) shows the fitting of ESR mode to extract the gyromagnetic ratio, $\gamma$. Overlaid plot in (b) shows the simultaneous fitting of acoustic and optical modes to the macrospin model defined by MacNeill \emph{et al.}\cite{Macneill_GigahertzCrCl3_2019}. (c) shows the change in $M_S$ with temperature at different angles.}
\end{figure*}

To extract various parameters for the two sublattice system, the resonance modes have been fit to a macrospin model presented by MacNeill \emph{et al.}\cite{Macneill_GigahertzCrCl3_2019}. This model gives a coupled Landau-Lifshitz-Gilbert equation for a two sublattice system. It can be written in the matrix form as

\begin{equation} \label{eq:2}
    \begin{pmatrix} {\omega^2}_a \big( H, \psi \big) - \omega^2 & \Delta^2 \big( H, \psi \big) \\ \Delta^2 \big( H, \psi \big) & {\omega^2}_o \big( H, \psi \big) - \omega^2 \end{pmatrix} = 0
\end{equation}
where $\omega_a = 2 \pi f_a$, $f_a$ is the frequency of the acoustic resonance mode, and $\omega_o = 2 \pi f_o$, $f_o$ is the frequency of the optical resonance mode. $\Delta = \mu_0 \gamma H ( \frac{2H_E} {2H_E + M_S} \sin^2 \psi \cos^2 \psi )^{1/4}$ is the magnon-magnon coupling. On calculating the eigenvalues of the matrix, we get two solutions with five dependent variables, DC magnetic field ($H$), interlayer exchange field ($H_E$), saturation magnetization ($M_S$), gyromagnetic ratio ($\gamma$) and angle with respect to $xy$-plane ($\psi$). As the angle is not changed during a particular measurement, $\psi$ is a fixed value. The $g$-factor for bulk CrCl$_3$ ($\approx 2$) has been determined previously\cite{chehab1991two}. $\gamma$ can also be confirmed by fitting a straight line to the electron spin resonance observed for temperatures greater than the ordering temperature of CrCl$_3$. Such a fit has been shown in Fig.~\ref{fig:fig8}(a) with $\gamma = 29.8~$GHz/T for temperature, $T = 20~$K. Thus, we have constrained the values of $\gamma$ to be between 28 GHz/T and 30 GHz/T. The frequency and field dependence of the modes have been visually determined and recorded using Engauge digitizer software. The two resonance modes are then simultaneously fit using Levenberg–Marquardt method to determine the values of $H_E$ and $M_S$. A sample fit has been shown is Fig.~\ref{fig:fig8}(b) for $T \approx 1.5$K and  $\psi \approx 25^\circ$. The variation of $M_S$ with temperature for different angles has been shown in Fig.~\ref{fig:fig8}(c).

\section{Fit to position of the spin-wave resonance modes}
\begin{figure*}[h]
\includegraphics{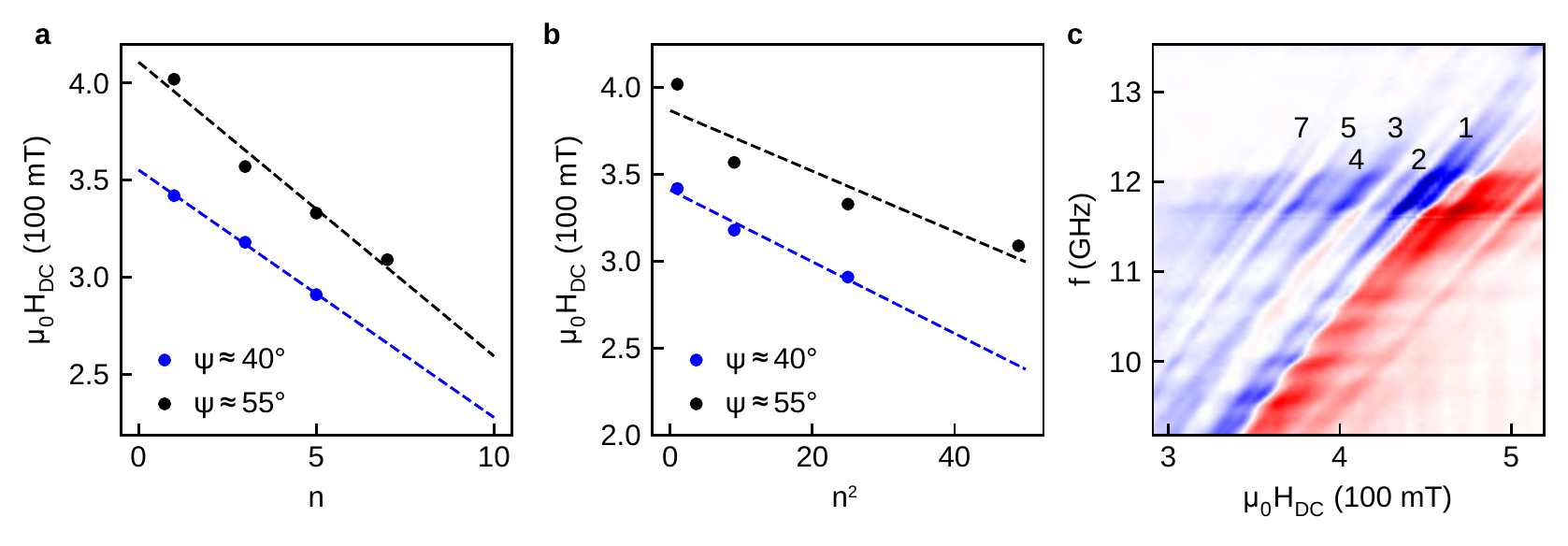}
\caption{\label{fig:fig9} \textbf{Comparison of spin-wave resonance mode fits for $n$ and $n^2$.} (a) and (b) shows the best fit for the linear dependence of the position of the spin-wave resonance modes with odd integers, $n$ and square of the odd integers, $n^2$ respectively. (c) shows the spin-wave resonance modes for $\psi \approx 55^\circ$. All the observed even and odd modes have been marked by their respective integer indices.}
\end{figure*}

For a uniform magnetization in a crystalline film, the Kittel model\cite{kittel_excitation_1958} gives dependence of the spin-wave resonance modes directly proportional to the square of an odd integer, $n^2$.
\begin{equation} \label{eq:3}
    H_n = H_0 -n^2\frac{D}{g\mu_B}\frac{\pi^2}{L^2}
\end{equation}
where $H_0$ is the position of the theoretical ferromagnetic resonance mode, $D$ is the exchange stiffness constant and $L$ is the thickness of the sample. However, for non-uniform magnetization, Portis' model\cite{portis1963low} gives a linear dependence with odd integer, $n$, given by equation~(\ref{eq:1}).

Fig.~\ref{fig:fig9}(a) and (b) show the best fit for the position of the spin-wave resonance modes with $n$ and $n^2$ respectively. While statistical determination of the best fit in both the cases is difficult for such a small data-set, we can use the $r^2$ from the linear regression analysis to compare the two fits. For $\psi \approx 55^\circ$, we find $r^2 = 0.972$ for linear dependence with $n$ and $r^2 = 0.863$ for linear dependence with $n^2$. Hence, we can conclude that the fit with linear dependence with $n$ is better. We also observe some weak signals between the spin-wave resonance modes corresponding to the odd integers, $n$, in regions close to a circuit resonance around 12 GHz. These weak signals may correspond to the even modes of the spin-wave resonances which can get excited in the presence of an inhomogenous field \cite{kittel_excitation_1958}. The even and odd modes have been marked in Fig.~\ref{fig:fig9}(c), observed in the case of $\psi \approx 55^\circ$

\section{Fitting details for magnon lifetime and Gilbert damping}
\begin{figure*}[h]
\includegraphics{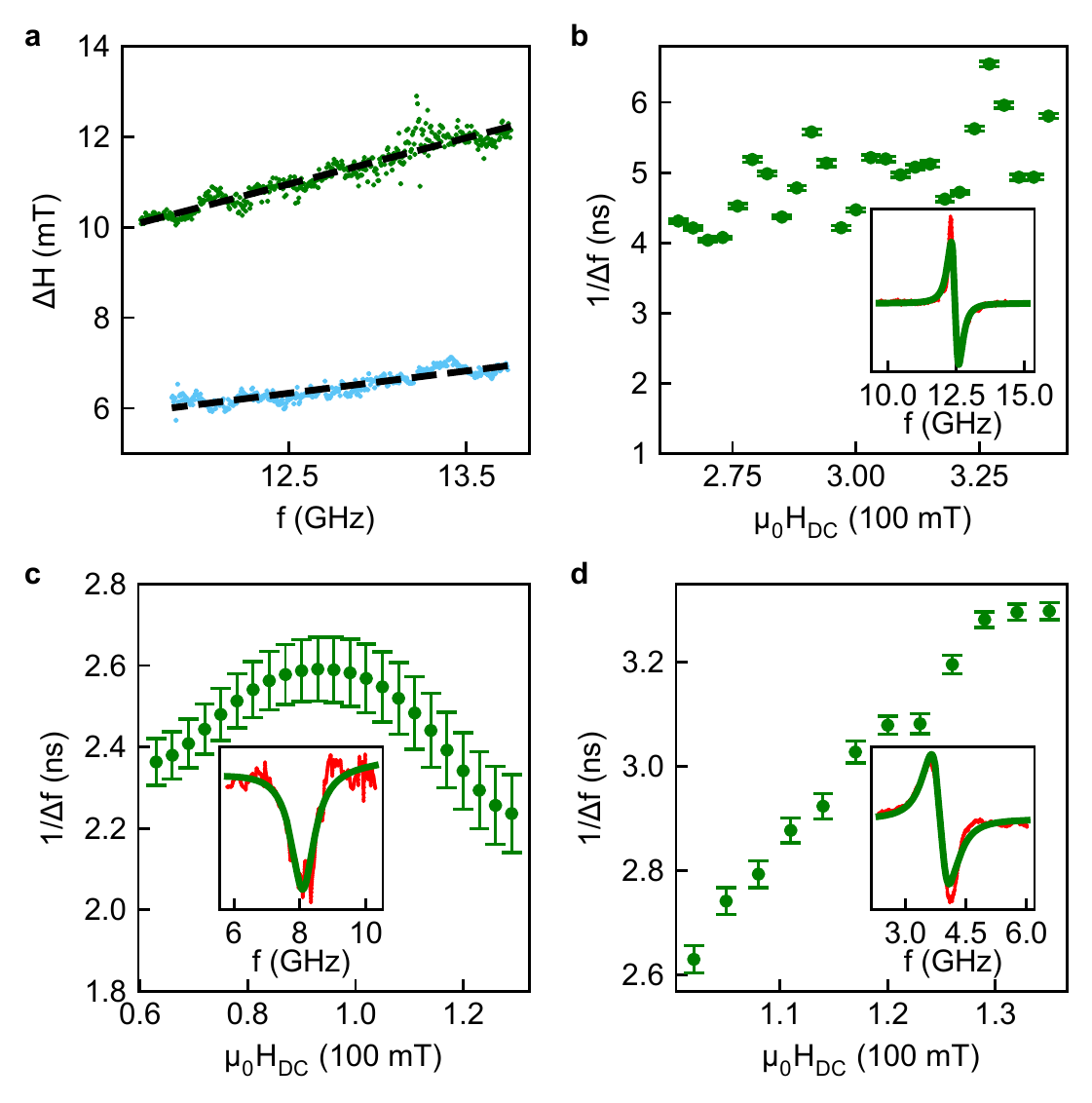}
\caption{\label{fig:fig10} \textbf{Fit to determine the Gilbert damping constant.} (a) shows the best fit for the linear dependence of the FWHM of the ferromgnetic resonance linewidth as a function of frequency. Green and blue points correspond to $\mathbf{H}_\textrm{DC}$ applied in-plane of CPW along $\hat{H}_x$ and $\hat{H}_y$ respectively. The calculated Gilbert damping coefficient are (a) $2.28\times10^{-3}$ and (b) $1.37\times10^{-3}$. (b),(c) and (d) show the magnon lifetime as a function of static magnetic field, $\mathbf{H}_\textrm{DC}$, applied in-plane along $\hat{H}_x$, for ferromagnetic resonance and optical and acoustic resonance modes respectively. Inset in (c) shows the fitting of the $|S_{21}|^2$ data with a Lorentzian while inset in (b) and (d) show the fitting of $\frac{1}{\mu_0}\frac{d}{dH_{\textrm{DC}}}|S_{21}|^2$ with a derivative of Lorentzian for better fitting.}
\end{figure*}

According to Polder susceptibility tensor, imaginary part of susceptibility ($\chi''$) for a ferromagnet is a Lorentzian function of frequency\cite{spindynamics}. Furthermore, $|S_{21}|^2$ is directly proportional to $\chi''$ where $|S_{21}|^2$ is the ratio of power of signal at port 2 with respect to port 1. Hence, $|S_{21}|^2$ takes the form of a Lorentzian function of frequency; inverse of the line-width or FWHM of this Lorentzian ($\Delta f$) gives the life-time of the ferromagnetic magnons to be around 5.0 ns, as shown in Fig.~\ref{fig:fig10}(b). In case of ferromagnetic resonance, $|S_{21}|^2$ takes the form of a Lorentzian function of magnetic field as well; the corresponding broadening is proportional to the frequency for the Gilbert damping contribution and is given by $\mu_0 \Delta H=\frac{4 \pi \alpha}{\gamma} f$, where $\alpha$ is the Gilbert damping parameter and $\gamma$ is the gyromagnetic ratio\cite{wang2018temperature,rossing1963resonance}. Thus from the slope of $\mu_0 \Delta H$ vs $f$ plot, we determine the Gilbert damping parameter for the ferromagnetic case to be around $2.0\times10^{-3}$. As seen in Fig.~\ref{fig:fig10}(a), variation of linewidth with frequency has been determined for in-plane static magnetic field along $\hat{H}_x$ and $\hat{H}_y$ (shown in green and blue color respectively). For the antiferromagnetic resonance, correspondence between frequency and field spaces is more complex, and we only calculate the lifetime of the antiferromagnetic magnons from the frequency-space Lorentzian fit. For the acoustic and the optical branches, the lifetimes are around 3.0 ns and 2.4 ns respectively at low fields. Fig.~\ref{fig:fig10}(c) and (d) shows the variation of the magnon lifetime with ${H}_\textrm{DC}$ for optical and acoustic resonance modes respectively. Fig.~\ref{fig:fig10}(b) shows the variation of the magnon lifetime for the ferromagnetic resonance. Sample fits for both the resonance modes are shown in inset of the respective plots. Our estimation of these parameters are approximate and extracted from parts of data which are less noisy. Further background elimination methods and noise optimization of measurement setup along with denser data-point spacing can be implemented for improvement of the extraction of these parameters.


\end{document}